\documentclass[preprint,10pt]{sigplanconf}
\usepackage{graphicx}
\usepackage{url}

\usepackage{amsmath}
\usepackage{listings}
\usepackage{algpseudocode}
\usepackage{algorithm}
\begin{document}
\newcommand{\code}[1]{\texttt{#1}}
\newcommand{\program}[1]{\texttt{#1}}
\newcommand{\comment}[1]{(\textit{#1})}
\include{rcj-defs}

\setlength{\pdfpageheight}{\paperheight}
\setlength{\pdfpagewidth}{\paperwidth}
\lstset{numbers=left,captionpos=b,basicstyle=\footnotesize,escapeinside={(*}{*)}}
\lstdefinelanguage{IrGL}[ISO]{C++}{morekeywords={ForAll,ReduceAndReturn,In,Atomic,Else,Exclusive,GlobalBarrierSync,Iterate,While,Any,Once,Invoke,Pipe,Kernel,Respawn,Retry}}
\conferenceinfo{CONF 'yy}{Month d--d, 20yy, City, ST, Country}
\copyrightyear{20yy}
\copyrightdata{978-1-nnnn-nnnn-n/yy/mm}
\doi{nnnnnnn.nnnnnnn}
\newcommand\IrGL[0]{IrGL}
\newcommand\lang[1]{\textsf{#1}}

\titlebanner{}        %
\preprintfooter{}   %

\title{Lowering IrGL to CUDA}
\subtitle{}

\authorinfo{Sreepathi Pai \and Keshav Pingali}
           {The University of Texas at Austin.}
           {sreepai@ices.utexas.edu,pingali@cs.utexas.edu}

\maketitle

\begin{abstract}
The IrGL intermediate representation is an explicitly parallel
representation for irregular programs that targets GPUs.  In this
report, we describe IrGL constructs, examples of their use and how
IrGL is compiled to CUDA by the Galois GPU compiler.

\end{abstract}


\section{Compiling \IrGL{}}
\label{sec:compiler}

Our implementation of the \IrGL{} compiler is written in Python and operates on an AST of \IrGL{} constructs~(Listing~\ref{lst:irglasdl}).
Apart from the constructs in Table~\ref{tbl:irglstmts}, this AST also contains the \lang{CBlock} construct for C++ code used in writing the operator.
Our compiler parses this code for well-formedness as well as generating read/write sets, but is limited to C99 syntax and hence also accepts annotations to describe read/write sets.
The compiler generates CUDA output, targeting Kepler and Maxwell GPUs.
In the remainder of this report, we describe how the \IrGL{} AST is lowered to CUDA. 
We assume a deep familiarity with CUDA. 
Listing~\ref{lst:irglasdl} contains the definition of this AST in ASDL~\cite{danielwang1997}.

We use the following typographical conventions in this document -- \lang{Terminals} in the AST  are indicated by a sans-serif font.
Attributes on AST nodes, as well as values in code are represented by \texttt{typewriter} font.

\begin{table*}
\begin{tabular}{|p{5cm}|p{12cm}|}
\hline
 \textbf{Construct} &  \textbf{Semantics}  \\ \hline

\multicolumn{2}{|c|}{\bf Kernel Constructs} \\ \hline

ForAll (\textit{iterator}) \{ \textit{stmts} \}  &
Traverse \textit{iterator} in parallel executing \textit{stmts} \\ \hline

ReduceAndReturn (\textit{bool-expr}) &
Reduce values of \textit{bool-expr} and return as kernel return value. The actual reduction, one of \lang{Any} or \lang{All}, is specified at kernel invocation. \\ \hline

Atomic (\textit{lock-expr}) \{ \textit{locked-stmts} \} [ Else \{ \textit{failed-stmts \} } ] &
Acquire \textit{lock-expr} and execute \textit{locked-stmts}. If an \lang{Else} block provided, execute \textit{failed-stmts} if \textit{lock-expr} was not acquired. Implements divergence-free blocking locks~\cite{ramamurthy2011}. \\ \hline

Exclusive (\textit{object, elements}) \{ \textit{locked-stmts} \} [ Else \{ \textit{failed-stmts \} } ] &
Try once to acquire locks for \textit{elements} in \textit{object} and execute \textit{locked-stmts} on succeeding. On failure execute \textit{failed-stmts} if provided otherwise execute next statement. One thread is guaranteed to execute \textit{locked-stmts} on conflicts. \\ \hline

SyncRunningThreads & Compiler-supported safe implementation of GPU-wide global barriers~\citep{xiao2010} \\ \hline

Retry \textit{item} (\textit{or} Respawn \textit{item}) & Push \textit{item} into a retry worklist and re-execute the kernel. Use of \lang{Retry} indicates a runtime conflict and triggers conflict management in the runtime (e.g. serial execution).  \\ \hline
\multicolumn{2}{|c|}{\bf Orchestration Constructs} \\ \hline

[Any \textbar{} All (] Invoke \textit{kernel}(\textit{args}) [)] &
Invoke \textit{kernel}, passing current worklists if kernel uses them.
 \\  \hline
Iterate [While \textbar{} Until Any \textbar{} All] \textit{kernel}(\textit{args}) [Initial (\textit{init-iter-expr})] &
Iteratively invoke \textit{kernel} until termination condition is met or worklist is depleted. if invoked standalone, establish fresh worklists using \textit{init-iter-expr} for initialization, else pass current worklists.
\\ \hline
Pipe [Once] \{ \textit{stmts} \} &
Establish worklists to be used by \textit{stmts}.
Without \lang{Once}, repeat \textit{stmts} until worklists are empty.
Nested \lang{Pipe}s will not establish worklists.
\\ \hline

\end{tabular}
\caption{Summary of \IrGL{} Statements, [] indicate optional parts, \textbar{} indicates options. See Listing~\ref{lst:irglasdl} for a formal definition of the abstract syntax tree.}
\label{tbl:irglstmts}
\end{table*}

\section{Overall Structure of the AST}

The \IrGL{} AST is rooted at the \lang{Module} node.
The only supported children of \lang{Module} are \lang{Kernel} and global-level declarations.
The \lang{Names} node provides support for importing foreign names (such as \texttt{\#define} constants) from C into the local \IrGL{} symbol table.
Many constructs in the AST, such as \lang{While}, \lang{If}, etc. only serve to expose control flow to the \IrGL{} compiler.
C code blocks, represented by \lang{CBlocks}, must observe single-entry, single-exit control-flow behaviour, akin to basic blocks, but are allowed to call functions.

Not shown in the AST definition are compiler-specific \textit{annotations} that can be applied to certain nodes.
For example, the CUDA \code{\_\_launch\_bounds\_\_} annotation conventionally used to indicate register-usage restrictions to the CUDA compiler is also supported by our \IrGL{} compiler (which passes it through, but see Section~\ref{sec:threads}).
Other significant annotations will be mentioned when the nodes they annotate are discussed below.

\section{Compiling Kernels}
\label{sec:compilingkernels}
A \lang{Kernel} node designates a \textit{plain} \IrGL{} kernel, a \textit{host} kernel (\code{host=true}) or a \textit{device} kernel (\code{device=true}).
Host kernels execute on the CPU.
Device kernels correspond directly to CUDA \code{\_\_device\_\_} kernels.

Only a host \lang{Kernel} may use the \lang{IrGL} orchestration constructs -- \lang{Invoke}, \lang{Iterate} and \lang{Pipe}.
Similarly, only a non-host, non-device \lang{Kernel} can use the \IrGL{} kernel constructs -- \lang{Atomic}, \lang{Exclusive}, \lang{Retry}, \lang{Respawn}, \lang{ReduceAndReturn} and \lang{ForAll}.
Host kernels may use \lang{ForAll}, but it is treated as \lang{For} by our current compiler.

The compiler primarily uses device kernels when implementing the iteration outlining optimization.
While user-provided device kernels are supported, they are treated as opaque and are largely ignored by our compiler.

The following subsections discuss how we compile the \lang{ForAll}, \lang{Atomic}, \lang{Exclusive} and \lang{ReduceAndReturn} kernel constructs. 
We defer discussion of \lang{Retry} and \lang{Respawn} to Section~\ref{sec:compilingorc}.

\subsection{Compiling ForAll}
The iterations of the outermost \lang{ForAll} in an \IrGL{} kernel are mapped to CUDA threads.
Each CUDA thread usually executes multiple iterations.
Section~\ref{sec:threads} describes the process the compiler uses to determine the number of CUDA threads to use for a kernel.

The object stored in \texttt{iterator} represents a random-access iterator and the order of iteration execution is not defined.
By default, consecutive iterations of the \lang{ForAll} are mapped to consecutive CUDA threads.
However, other mappings are possible and these mappings are represented by compiler-specific annotations.
For example, when implementing the Retry Backoff optimization, a mapping that distributes contiguous blocks of iterations to a single thread is used to reduce conflicts.

\lang{For} is used to represent a loop whose iterations cannot be executed in parallel. 
When the amount of parallelism can only be discovered at runtime, as in most irregular graph algorithms, \lang{ForAll} is used with additional synchronization inside the body of the loop.

\subsection{Compiling Atomic and Exclusive}
\IrGL{} provides two statements -- \lang{Atomic} and \lang{Exclusive} -- that allow iterations of a \lang{ForAll} loop to implement mutual exclusion.
Both these statements implement functionality that are hard to get right~\cite{ramamurthy2011, alglave2015}.
\lang{Atomic} and \lang{Exclusive} currently use software implementations but can be recompiled to use proposed hardware primitives~\cite{ramamurthy2011,wwlfung2012} if such primitives become available.

\subsubsection{Atomic}
\lang{Atomic} implements an atomic section, a block of code that is executed under control of a single lock.
\lang{Atomic} can be nested.
Two forms of \lang{Atomic} are supported, a default blocking form that waits for the lock to be acquired.
The other form, indicated by a non-empty \texttt{fail\_stmts} (i.e. \lang{Else}), is non-blocking and executes the statements in \texttt{fail\_stmts} when the lock was not acquired.

\lang{Atomic} provides as a safe alternative to spinlocks, since spinlocks can deadlock on GPUs due to warp divergence.
Internally, \lang{Atomic} uses \code{atomicCAS} to set the state of a lock variable.
If the \code{atomicCAS} fails to acquire the lock and the \lang{Atomic} is blocking, a divergence-safe loop similar to that described in~\cite{ramamurthy2011,nyland2013} is generated by the compiler to reattempt locking.

We illustrate the use of \lang{Atomic} using Bor\r{u}vka's algorithm for minimum spanning tree.
Bor\r{u}vka's algorithm begins by treating each node of the input graph as a {\it component}. Then, it finds the minimum cross-component edge out of each components.
These edges are added to the minimum spanning tree, and the components they connect are merged (or {\it unified}). The procedure then repeats on these merged components until only one component remains or no cross-component edges can be found (i.e. in a disconnected graph).

Particularly challenging is the implementation of finding the minimum edge out of a component. Since a component can consist of many nodes, recording the minimum edge requires at least two updates that must be carried out atomically -- the minimum weight and the edge itself. No CUDA primitive suffices to perform multiple updates atomically. Previous implementations, notably that of \citet{vineet2009}, store the weight and edge identifier as bitfields of a 32-bit integer, which would allow use of a single \code{atomicMin} at the cost of severely limiting the generality of the resulting code and its applicability to input graphs.

Listing~\ref{lst:boruvkamst} describes the \code{find-min-edge} kernel in the \IrGL{} implementation that uses an \lang{Atomic} to update the component's data in an atomic context.
Note that this instance of \lang{Atomic} is a blocking lock (i.e. no \lang{Else} clause), so it can use a ticket lock from our runtime by simply setting a compiler flag. 

\begin{lstlisting}[language=IrGL,float,caption={Find-Minimum-Cross-Component-Edge kernel of Bor\r{u}vka's MST algorithm. The worklist initially contains all nodes.}, label=lst:boruvkamst]
ForAll(nidx In wl) {
  n = wl.pop(nidx);
  n_component = components[n];
  minwt = INF;

  for(e In edges) {
    // find minimum cross-component edge
    // out of node n; store weight in minwt,
    // and edge id in minedge
  }

  Atomic(component_locks[n_component]) {
    if(component_minwt[n_component] > minwt) {
      component_minwt[n_component] = minwt
      component_minedge[n_component] = minedge
      // other updates
    }
  }

  if(node has cross-component edge) {
    wl.push(n)
  }
}
\end{lstlisting}

\subsection{Exclusive}
\lang{Exclusive} encloses a block of code that must acquire a large number of locks.
Internally, threads are assigned priorities so that at least one thread always acquires all the locks it needs.
\lang{Exclusive} never blocks and may not be nested.
The statements in \code{fail\_stmt} are executed if the locks were not acquired.

The set of locks to be acquired is obtained from an array of lock indices. 
In the simplest form, \texttt{objs\_lock} indicates an \lang{Array} which contains the lock indices to be acquired.
In the \lang{ArrayIterator} form, \texttt{objs\_lock} specifies an array iterator that yields the indices to be locked.

Consider the use of \lang{Exclusive} in Delaunay Mesh Refinement~(DMR).
For DMR, the key kernel is \code{refine} which, when presented with a worklist of bad triangles, fixes each one of them in parallel. Each thread must have exclusive access to the triangles in the cavity of its bad triangle, as well as to the triangles that form the boundary of the cavity. The \lang{Exclusive} construct is key to simplifying the implementation of DMR. Listing~\ref{lst:dmr} illustrates how the triangles in the cavity are passed as input to \lang{Exclusive} which then permits access to triangles in the cavity to one thread.

\begin{lstlisting}[language=IrGL,float,caption={Simplified Refine kernel in Delaunay Mesh Refinement.}, label=lst:dmr]
ForAll(btidx In wl) {
  bad_triangle = wl.pop(btidx);

  build_cavity(bad_triangle, &cavity_size, &cavity);

  Exclusive(mesh, cavity_size, cavity) {
    delete_cavity(...)
    // create new triangles
  }

  SyncRunningThreads();
}
\end{lstlisting}

The \lang{Exclusive} statement is implemented using a three-phase algorithm, with each phase separated by \lang{SyncRunningThreads}.
Our implementation is similar to the \textit{race--prioritycheck--check} scheme described in~\cite{nasre2013morph}.
In the first phase, \lang{Exclusive} claims the locks supplied for the executing thread.
If multiple threads claim the same lock, only one claim is allowed.
In the second phase, each thread checks to see if its claim for every lock stands. 
If another thread was granted the claim, then the threads that lost the claim attempt to win priority over the claim.
In the third phase, each thread checks to see if it still retains the claims it sought.
Any threads that do so proceed to execute the statements within the \lang{Exclusive}, while those that do not move to the next statement or execute the \lang{Else} clause if one is supplied.
Our use of \lang{SyncRunningThreads} for implementing \lang{Exclusive} places restrictions on its use.
An \lang{Exclusive} must be placed in a location that will be uniformly executed by all threads and thus cannot be nested.
In practice, this means that \lang{Exclusive} may only be directly placed underneath the outermost \lang{ForAll}.

\subsection{Compiling SyncRunningThreads}
\label{sec:syncrunningthreads}

Like CPUs, GPUs can create and execute many more threads than can run concurrently on hardware.
However, unlike CPUs, a GPU thread usually runs to completion and cannot be preempted.
Thus, the notion of a global barrier that synchronizes all threads does not readily translate to the GPU.
If all threads being synchronized are not running concurrently, the global barrier will deadlock.

Nevertheless, barrier-like functionality is useful, even if it is limited to only those threads that are running concurrently.
Such ``device-wide barriers'', have been described previously~\cite{xiao2010} and are supported in \IrGL{} through the \lang{SyncRunningThreads} statement, though our implementation is derived from code used in~\cite{merrill2012}.
Safe use of \lang{SyncRunningThreads} requires that a GPU kernel never be launched with more physical threads than can run concurrently.
This number can vary from GPU to GPU and also depends on the size of the CUDA thread block.

CUDA~6.5 introduces the occupancy API that allows this number to be calculated at runtime for a kernel for each GPU present in the system.
When our compiler generates code to launch an \IrGL{} kernel that uses \lang{SyncRunningThreads}, it limits the number of threads using this occupancy API to ensure deadlock-free execution.
We note this method is not portable to other devices~\cite{sorensen2016iowcl} and even on NVIDIA GPUs, assumes that all thread blocks of the kernel will eventually execute concurrently. 

\subsection{Compiling ReduceAndReturn}

The \lang{ReduceAndReturn} statement is used to construct a return value for an \IrGL{} kernel using a reduction.
The actual reduction is specified when invoking the kernel using \lang{Invoke} or \lang{Iterate}.
Our compiler currently supports the \lang{Any} and \lang{All} reductions, and therefore the value to be reduced is a boolean expression stored in \texttt{value}.

\lang{Any} returns \code{true} if any \code{value} evaluated to \code{true}.
\lang{All} returns \code{true} only if all \code{value} evaluated to \code{true}.

\lang{ReduceAndReturn} terminates execution of the kernel, \textit{except} when invoked inside a \lang{ForAll} when it only terminates the current iteration.

The simplest compilation of \lang{ReduceAndReturn} uses global memory storage and CUDA atomic instructions to implement these reductions.
However, this can be made cheaper by re-using the cooperative conversion optimization machinery.
In effect, each CUDA thread partially aggregates the \lang{ReduceAndReturn} values, with atomics being used only to aggregate the values of each CUDA thread block.
Unfortunately, since CUDA does not support virtual functions, we must generate multiple variants for each kernel (e.g. by using C++ templates) for each reduction used in the calling \lang{Invoke} or \lang{Iterate}.

\section{Compiling Orchestration}
\label{sec:compilingorc}

The orchestration constructs \lang{Invoke}, \lang{Iterate} and \lang{Pipe} are used to invoke \IrGL{} kernels.
Since data-driven \IrGL{} kernels often use worklists, the \lang{Iterate} and \lang{Pipe} also setup worklists.
They also execute a series of \IrGL{} kernels until the worklist is empty since iterative exection is a common pattern.

\IrGL{} provides a default worklist object named \code{WL} that exposes \code{push}, \code{pop} and an iterator to each kernel. 
Therefore, \code{pop} and \code{push} are encoded in the AST as \lang{MethodInvocation} on this object and do not appear as first-class AST nodes.

\subsection{Worklist Mechanics}

\begin{lstlisting}[language=IrGL,float,caption={Level-by-level BFS kernel using a worklist},label=lst:bfs]
Kernel BFS(graph, LEVEL) {
  ForAll(wlidx In wl) {
    n = wl.pop(wlidx)
    ForAll(e In graph.edges(n)) {
      if(e.dst.level == INF) {
        e.dst.level = LEVEL;
        wl.push(e.dst.id);
      }
    }
}

LEVEL=0
Iterate BFS(graph, LEVEL) Initial [src] {
  LEVEL++;
};
\end{lstlisting}

Kernels use worklists to manage work and as a means of communication of work between kernels. \IrGL{} provides a default worklist to every kernel.
A kernel may \code{push} values onto a worklist to enqueue work, and may \code{pop} values off the worklist to perform work, usually using a \lang{ForAll}. 
\IrGL{} worklists exhibit bulk-synchronous behaviour -- work items pushed during an invocation cannot be popped in the same invocation.

Worklists are created and managed by \lang{Iterate} and \lang{Pipe} constructs.
\lang{Iterate} is best illustrated by the BFS code in Listing~\ref{lst:bfs}.
It creates a worklist, initially populated with $src$ and invokes the \code{BFS} kernel repeatedly until the worklist is depleted.
After every invocation, code in \texttt{stmts} is executed. In this example, the \code{LEVEL} variable is incremented.
The automatically created worklist is not available beyond the execution of the \lang{Iterate} statement.

\begin{lstlisting}[language=IrGL,float,caption={Example of Pipe in DMR},label=lst:dmrpipe]
Pipe Once {
  Invoke identify_bad_triangles(mesh);
  printf(``initial bad: %

  // looping Pipe
  Pipe {
    Invoke refine(mesh);
    ... // other mesh maintenance code

    // only among newly created triangles
    Invoke incremental_id_bad_triangles(mesh);
  }

  // sanity check
  Invoke identify_bad_triangles(mesh);
  printf(``final bad: %
}
\end{lstlisting}

The \lang{Pipe} statement establishes a shared worklist for the \lang{Iterate}, \lang{Invoke} and \lang{Pipe} statements within it.
A \lang{Pipe} may execute once or loop until the worklist is empty.
Inside a \lang{Pipe}, the values pushed by an invocation of a kernel are forwarded to the next kernel in the pipe. Listing~\ref{lst:dmrpipe} illustrates the use of \lang{Pipe} in the main loop of the DMR benchmark. 
After receiving an initial set of bad triangles, the inner \lang{Pipe} iteratively refines the mesh, communicating the worklists between the two kernels inside the pipe.

The ``flow'' of worklists between kernels is not fixed at compile time. 
For example, Listing~\ref{lst:dynamicpipe} is perfectly valid \IrGL{} code.
Depending on what \code{cond} evaluates to, the worklist produced by \code{A} may be consumed by either \code{B} or \code{C}.

\begin{lstlisting}[language=IrGL,float,caption={Example of Dynamic Piping},label=lst:dynamicpipe]
Pipe Once {
  Invoke A();

  if(cond) {
    Invoke B();
  else {
    Invoke C();
  }
}
\end{lstlisting}

\subsection{Compiling Pipe}

The actual creation and communication of worklists between kernels is the responsibility of the \lang{Pipe}/\lang{Iterate} statements.
Currently, the outermost \lang{Pipe} (or \lang{Iterate}) statement in a host \lang{Kernel} creates a \textit{pipe context}.
All nested \lang{Pipe}, \lang{Iterate} and \lang{Invoke} statements inherit this pipe context.
In our implementation, the pipe context contains the incoming, outgoing and retry worklists named \code{in}, \code{out} and \code{retry} respectively.
The \code{wlinit} attribute specifies the size of the worklist (\texttt{size}) and how the initial worklist is populated, which is implementation-dependent.
For example, our compiler supports initializing worklists from a list of scalar expressions (\lang{WorklistInitializer}) or from an array (\lang{WorklistInitializerFromArray}).

When compiling an invocation to a kernel that reads, writes or iterates over the \code{WL} object, all \code{pop}s are executed on the \code{in} worklist. 
Similarly all \code{pushes} are executed on the \code{out} worklist, using cooperative conversion where applicable to improve performance.
Workitems to be retried are pushed into the \code{retry} worklist.

\begin{figure}
\centering
\includegraphics[scale=0.5]{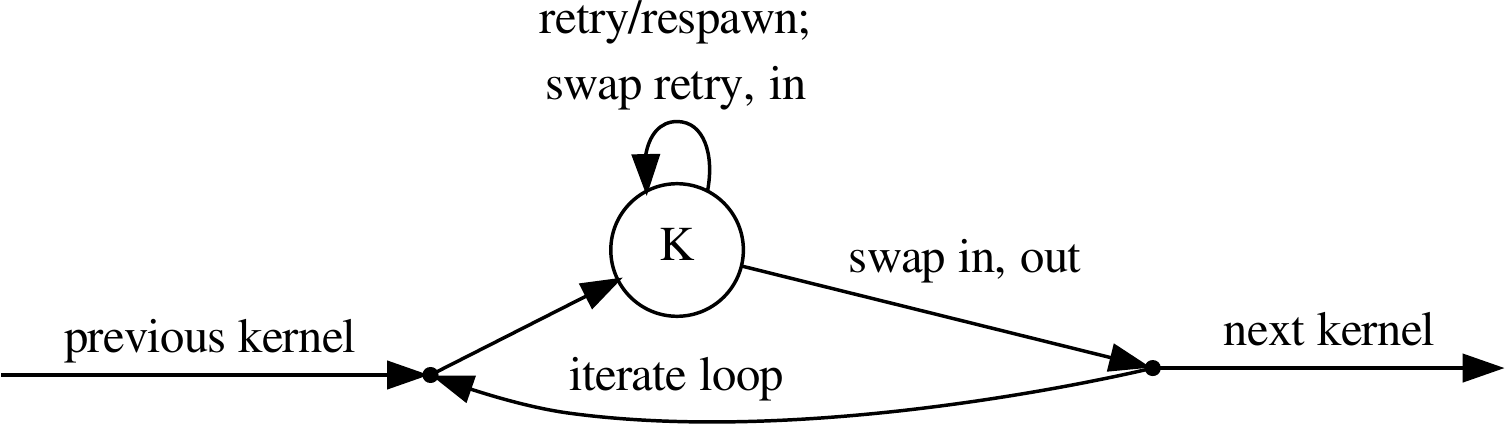}
\caption{Flow control and worklist management for \lang{Iterate}}
\label{fig:iterateflow}
\end{figure}

\begin{figure}
\centering
\includegraphics[scale=0.5]{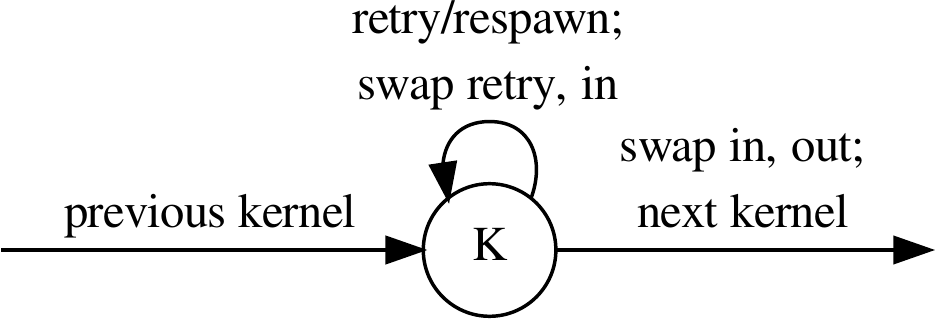}
\caption{Flow control and worklist management for \lang{Invoke} on a kernel that uses worklists}
\label{fig:invokeflow}
\end{figure}

Figures~\ref{fig:iterateflow} and \ref{fig:invokeflow} illustrate how control flows within a \lang{Iterate} or \lang{Invoke}.
In general, flow is linear from the previous kernel to the next unless the kernel uses \lang{Iterate} in which case the kernel is invoked repeatedly until no more items are left to process.
Each invocation swaps the \code{in} and \code{out} worklists.
If the kernel uses the \code{retry} worklist, it will be invoked repeatedly, but the \code{in} and \code{retry} worklists are swapped, while the \code{out} worklist remains the same.

Our compiler also sets up storage to store the return value when compiling \lang{Invoke} and \lang{Iterate} for kernels that use \lang{ReduceAndReturn}.

Note that \lang{Iterate} is syntactic sugar for a loop that wraps \lang{Invoke} for kernels that do not use worklists.
\lang{Iterate} is essentially equivalent to a \lang{Pipe} for kernels that do use worklists.
Apart from terminating when the worklist is empty, it is possible to specify (in \texttt{extra\_cond}) additional conditions that will cause the loop to exit even when the worklist is not empty.
This extra condition may be combined with the empty worklist check using either \lang{And} or \lang{Or}.

\subsection{Compiling Kernel Invocations}
\label{sec:threads}

In the most general case, when the statement invoking the kernel, an \lang{Iterate} or \lang{Invoke} or \lang{Pipe}, lies in a host kernel, a kernel invocation compiles down to a CUDA kernel launch.
However, when the kernel invocation lies in the \textit{control} kernel of an outlined \lang{Pipe}, which is a CUDA \texttt{\_\_global\_\_}, it is compiled to a device kernel function invocation.
Our compiler also supports the use of CUDA Dynamic Parallelism when launching kernels from control kernels, but the performance is poor, and it is not recommended.

Since \IrGL{} kernels have no notion of threads, our compiler must also choose appropriate grid and thread block sizes for the CUDA launch.
If \lang{SyncRunningThreads} or \lang{Exclusive} are not used in the kernel, then any grid size can be used, with our compiler using a fixed grid size calculated from the number of multiprocessors in the GPU.
The use of these constructs requires that the grid size be chosen carefully as described earlier in Section~\ref{sec:syncrunningthreads}.
Without optimizations enabled, \IrGL{} kernels also naturally compile down to \textit{elastic} kernels~\cite{pai2013} and so can run with any thread block size.
However, when optimizations are enabled, the thread block sizes for a kernel may be constrained as we describe below.

Our compiler allows programmers to use the CUDA \texttt{\_\_launchbounds\_\_(maxthreadsperblock, minblocks)} annotation on individual kernels.
The optional \texttt{minblocks} parameter is advisory and requests the compiler to achieve a residency of at least \texttt{minblocks} on each multiprocessor of the GPU.
It is ignored by our compiler.
The \texttt{max\-threads\-per\-block} parameter, on the other hand, informs the CUDA compiler that the kernel will not be launched with more than \texttt{maxthreadsperblock} which changes the behaviour of the register allocator.
Attempting to launch a kernel with more than \texttt{max}\texttt{threads}\texttt{per}\texttt{block} will result in failure.
Thus, \texttt{\_\_launchbounds\_\_} establishes an upper bound on the thread block size that can be used by our compiler.

Using nested parallelism or cooperative conversion also imposes a constraint on the thread block size that can be selected for a kernel.
Essentially, both these optimizations make use of CUDA shared memory for communication with the size of shared memory used depending on the thread block size.
Similarly, some libraries that we use internally use C++ template parameters to specialize for a statically specified thread block size.
Such kernels are therefore limited to a fixed thread block size.

To summarize, \lang{IrGL} kernels can fall into three categories depending on the thread block size they support.
First are the \textit{ElasticBlock} kernels, which can execute with any thread block size.
Second are the \textit{ShrinkableBlock} kernels, which place an upper bound on their thread block size.
Finally, in the third category are the \textit{FixedBlock} kernels, which can only execute with a fixed thread block size.

If iteration outlining is not used, the constraint on one kernel does not affect another kernel.
However, when a  \lang{Pipe} containing different kernels is outlined to the GPU and dynamic parallelism is \textit{not} used, the thread block size chosen for the control kernel must satisfy the constraints on all kernels in the \lang{Pipe}.

Since the maximum thread block size is limited by CUDA to 1024 on all GPUs we support, the set of possible thread block sizes for a kernel $k$, denoted by $T_k$, is finite.
If $K$ is the set of kernels in a \lang{Pipe}, then $T_{control}$ is simply:

\begin{equation*}
T_{control} = \bigcap_{k \in K}T_{k}
\end{equation*}

Thus, the thread block size of the control kernel is simply the intersection of the domain sets for the constraint variables of each kernel.
If this intersection is empty, then iteration outlining cannot be performed on this \lang{Pipe}.
If this intersection contains multiple values, our compiler chooses the highest value.

It is possible for ElasticBlock and ShrinkableBlock kernels to support different thread block sizes in different \lang{Pipe}s.
However, for simplicity, our compiler picks a single thread block size for each kernel that is used at every invocation.

\section{Conclusion}

In this report, we have described the AST for IrGL and how it is lowered to CUDA.
Our scope has been limited to the primary \IrGL{} constructs since our intent was to provide a high-level overview of the process.
We hope that this document will also be helpful to understand the organization of the \IrGL{} compiler source that will be released separately.
Among this document's omissions, we note absence of a discussion regarding the annotations supported by our compiler as well as its support for selecting optimizations at the \lang{Block} level that allows a richer search space for auto-tuning, since these constructs are currently in flux.

\begin{lstlisting}[float=*,caption={IrGL Abstract Syntax Tree in ASDL},label=lst:irglasdl]
   module = Module(mod_stmts stmts)
   mod_stmts = Kernel(identifier name, param *params, 
   	       	      bstmts stmts, str ret_type, 
		      bool host, bool device) 
	     | CDeclGlobal(cdecl *decls, bool parse, bool dont_move)
	     | CBlock(str *stmts, bool parse)
	     | NOP()
	     | Names(str *names)

   param = (str type, identifier name)
   cdecl = (str type, identifier name, str initializer)

   bstmts = Block(stmts *stmts)
 
   stmts = Block(stmts *stmts)
         | Assign(str lhs, str rhs)
         | For(identifier ndxvar, object iterator, bstmts *stmts)
	 | ForAll(identifier ndxvar, object iterator, bstmts *stmts)
         | CFor(str init, str cond, str update, bstmts *stmts)

	 | While(str cond, bstmts *stmts)
	 | DoWhile(str cond, bstmts *stmts)

	 | Atomic(str lock, str lockndx, stmts *stmts, bstmts *fail_stmts)
	 | Exclusive(object_list objs_lock, stmts *stmts, bstmts *fail_stmts)

	 | Retry(str args, bool merge)
	 | Respawn(str args)
	 | CBlock(str *stmts, bool parse)
	 | CDecl(cdecl *decls, bool parse, bool dont_move) 

	 | If(str cond, bstmts *true_stmts, bstmts *false_stmts)
	 
	 | Invoke(aggr_func aggr, identifier kernel, str *args)
	 | Iterate(term_cond cond, aggr_func aggr, identifier kernel, 
	           str *args, worklist_initializer wlinit, bstmts *smts, 
		   extra_cond extra_cond)
	 | ReduceAndReturn(str value)
	 
	 | SyncRunningThreads()
	 | LocalBarrier()

	 | Pipe(bstmts *stmts, worklist_initializer wlinit, bool once)

	 | Expr(expr e)

   expr = CExpr(str expr, bool parse) 
        | MethodInvocation(str obj, identifier method, str obj_type, str *args)

   object_list = Tuple(identifier object, item_list *items)

   item_list = Array(int size, identifier name) 
             | ArrayIterator(identifier array, str start, str end, str step)

   term_cond  = While | Until
   aggr_func  = Any | All
   extra_cond = And(str cond) | Or(str cond)
   
   worklist_initializer = WorklistInitializer(str size, str *initial)
   			| WorklistInitializerFromArray(str size, identifier array, str array_size)

\end{lstlisting}

\bibliographystyle{abbrvnat}

\bibliography{irgl2cuda}

\end{document}